# Single-atom electron paramagnetic resonance in a scanning tunneling microscope driven by a radiofrequency antenna at 4 K

T. S. Seifert[1], S. Kovarik[1], C. Nistor[1], L. Persichetti[1], S. Stepanow[1], P. Gambardella[1]

[1]Department of Materials, ETH Zurich, 8093 Zurich, Switzerland

Combining electron paramagnetic resonance (EPR) with scanning tunneling microscopy (STM) enables detailed insight into the interactions and magnetic properties of single atoms on surfaces. A requirement for EPR-STM is the efficient coupling of microwave excitations to the tunnel junction. Here, we achieve a coupling efficiency of the order of unity by using a radiofrequency antenna placed parallel to the STM tip, which we interpret using a simple capacitive-coupling model. We further demonstrate the possibility to perform EPR-STM routinely above 4 K using amplitude as well as frequency modulation of the radiofrequency excitation. We directly compare different acquisition modes on hydrogenated Ti atoms and highlight the advantages of frequency and magnetic field sweeps as well as amplitude and frequency modulation in order to maximize the EPR signal. The possibility to tune the microwave-excitation scheme and to perform EPR-STM at relatively high temperature and high power opens this technique to a broad range of experiments, ranging from pulsed EPR spectroscopy to coherent spin manipulation of single atom ensembles.

## I. INTRODUCTION

Scanning tunneling microscopy (STM) is a unique technique to achieve subatomic spatial resolution with simultaneous local spectroscopic information [1]. The demonstration of spin sensitivity in STM experiments [2-4] enabled the study of single magnetic atoms on a surface and their interactions [5-8]. Despite these great advances, the energy resolution remains limited in tunneling-spectroscopy modes by the thermal energy broadening of the electronic tip and sample states (>1 meV at 4 K). This broadening limits the precise sensing of low-energy excitations, e.g., spin-flip excitations, which motivated efforts to reduce the STM operational temperature to the mK range [9-12] and to apply large magnetic fields to obtain the required sensitivity [13].

Another promising way to overcome the thermally-limited energy resolution is to employ a resonance technique. In this regard, electron paramagnetic resonance (EPR) is an established method [14] that has found diverse applications such as the identification of free radicals in chemical reactions [15], detection of spin-labeled molecules in biological systems [16], or the study of molecular nanomagnets [17]. Following early attempts [18,19], Baumann et al. [20] were the first to convincingly demonstrate EPR of single atoms on a surface using STM. In these experiments [20-22], the authors studied single Fe, Ti, and Cu atoms on a thin insulating MgO layer grown on Ag(100) (see Fig. 1a).

In EPR-STM experiments, an external magnetic field $B_{\text{ext}}$ splits the energy of the states of the magnetic atom under investigation. A resonant microwave voltage that is fed through the tip wire to the STM tunnel junction induces transitions between these Zeeman-split states. Upon resonance, the spin-dependent conductivity of the tunnel junction varies, which is sensed by a spin-polarized STM tip through a magneto-resistive effect [20,23,24].

Although the mechanism underlying the EPR excitation is still under debate [25-28], EPR-STM has proven capable of measuring EPR

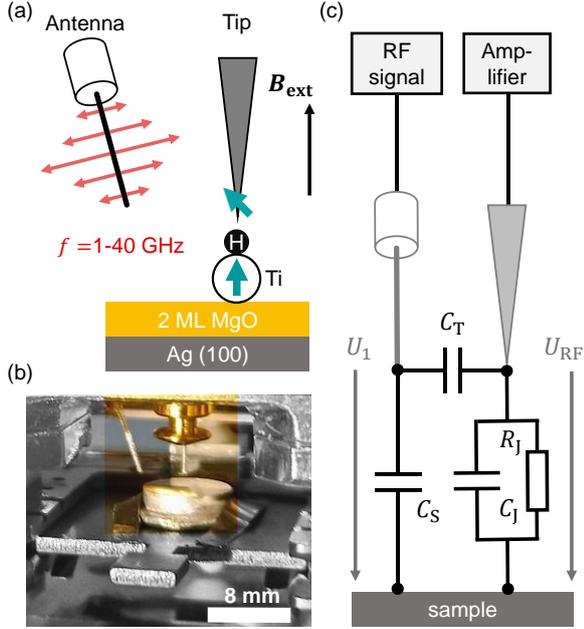

FIG. 1. Schematic of the EPR-STM setup. (a) A radiofrequency (RF) antenna is capacitively coupled to the STM tip. The resulting RF voltage $U_{RF}$ between tip and sample drives EPR of a spin-1/2 system (hydrogenated Ti) deposited on top of two monolayers (ML) of MgO on Ag(100). The external magnetic field $B_{ext}$ is applied parallel to the tip. Cyan arrows depict magnetic moments. (b) Photograph of the RF antenna used to couple RF voltages to the STM tunnel junction. For clarity, the photograph is rotated upside-down with respect to the actual STM geometry. (c) Equivalent circuit of the indirect RF coupling scheme including the capacitances between RF antenna and sample $C_S$, between RF antenna and STM tip $C_T$, and the tunnel-junction capacitance $C_J$ as well as the tunnel resistances $R_J$. The ratio between $U_{RF}$ and the RF voltage at the RF antenna $U_1$ determines the antenna efficiency. An RF generator provides RF excitation signals. The STM tip is connected to a transimpedance amplifier.

spectra of single atoms [20], probing dipolar and exchange interactions between different atoms on a surface [21,24,29-31], and the hyperfine interaction of isotopic species with a finite nuclear moment [22,32]. These results were obtained at temperatures lower than 1.2 K with rare exceptions at temperatures of up to 4 K, which, however, resulted in a drastically reduced signal-to-noise ratio [33,34].

Given the impressive results listed above, it is highly desirable to make EPR-STM accessible to a broader range of experiments. Crucial steps in this direction are maximizing the signal-to-noise ratio in EPR-STM and demonstrating routine operation at liquid-helium temperature and above. To achieve the first goal, the EPR excitation needs to be tailored to reduce the noise and maximize the EPR signal, which involves an optimized coupling of the radiofrequency (RF) voltage to the tunnel junction. Moreover, the efficiency of different EPR detection modes such as frequency and magnetic field sweeps, as well as amplitude, and frequency modulation of the RF excitation should be compared. Eventually, with stronger EPR excitation, the Rabi rate (i.e., the rate, at which the driven system undergoes population inversion) could overcome the spin-decoherence rate [33]. This strong excitation might thus enable coherent spin manipulations with EPR-STM, opening the way to performing quantum computation experiments with single atoms on surfaces [35].

In previous work, EPR-STM was achieved by amplitude modulation of the RF excitation. Furthermore, the RF voltage required to drive EPR was fed directly through the STM tip into the tunnel junction by combining the RF with the DC bias voltage outside the STM cryostat using a bias-Tee [34,36,37]. This approach limits the RF transmission efficiency [34,36], the signal-to-noise ratio, and involves complex modifications of the STM wiring.

Here, we implement and characterize an indirect RF coupling via an RF antenna close to the STM tip (see Fig. 1) [38,39] that results in significantly higher RF voltages across the tunnel junction than reported so far for frequencies up to 40 GHz. Further, we analyze the RF-coupling scheme in the tunnel junction and show that the RF antenna reaches unexpectedly high coupling efficiencies, of the order of unity, which we explain using a simple equivalent circuit model. In the second part, we show that our indirect RF-coupling scheme can drive EPR of single hydrogenated Ti atoms on a surface using a broad range of power (see Fig. 1). Remarkably, even at temperatures above 4 K an energy resolution of 1 µeV can be achieved. A comparison of different EPR-STM modes (sweep of radiofrequency or external field $B_{ext}$) highlights the advantage of sweeping $B_{ext}$, which significantly reduces the acquisition time of an EPR spectrum. We find that both schemes yield consistent magnetic moments of the investigated species. As far as we know, this is the first side-by-side comparison of these sweep modes with the same STM microtip and the same EPR species. Finally, we extend the repertoire of EPR-STM by exploring and comparing directly three RF-modulation schemes, namely, amplitude and frequency modulation. We point out how the right choice of modulation can further maximize the signal-to-noise ratio.

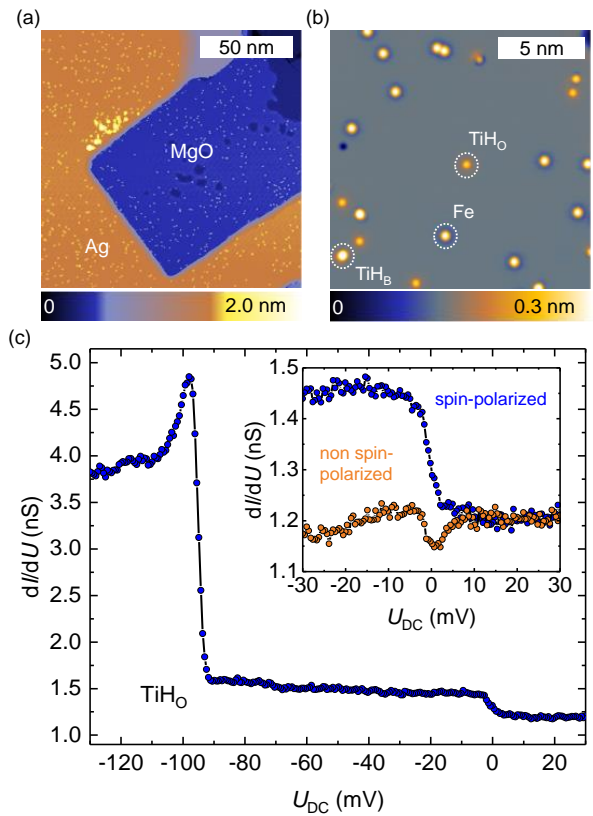

FIG. 2. Single magnetic atoms on MgO/Ag. (a) Large-scale constant-current image of a double-layer MgO island grown on a Ag(100) surface (DC bias $U_{DC} = 30$ mV, setpoint current 50 pA, temperature 4.5 K). (b) Detailed constant-current image of single Fe and hydrogenated Ti atoms (TiH, subscripts O and B refer to different binding sites) on a double-layer MgO island [same settings as in (a)]. (c) d$I$/d$U$ spectra recorded on top of a TiH$_O$ at 0.5 T showing spin contrast (feedback loop opened at setpoint current 50 pA, DC bias 30 mV). The inset of (c) shows detailed d$I$/d$U$ spectra on the same TiH$_O$ atom showing an STM microtip with and without spin contrast visible near zero DC bias.

## II. EXPERIMENT

### A. Setup

All experiments were performed with a Joule-Thomson STM (Specs GmbH) equipped with a superconducting magnet having a maximum out-of-plane magnetic field of 3 T. To upgrade the STM for EPR capabilities, we installed an RF-transmission line consisting of a semi-rigid coaxial cable (SC-119/50-SB-B with K-type connectors assembled by Coax Co., Ltd, length of 2.5 m) going from 300 K to the bottom of the He vessel and a flexible coaxial cable (Part No. 1070551 from Elspec, SMPM connector on one side, length of 0.3 m) going from the liquid-He vessel to the STM head. The RF cables were thermally anchored at different positions in the cryostat, such that no significant change of the liquid-He hold time was detected after the upgrade. The RF antenna forms the final part of the RF-transmission line. The antenna is made from the 5-mm-long unshielded inner conductor of the flexible coaxial cable, which is positioned as closely (~5 mm away) and as parallel (angle of ~30°) as possible to the STM tip (see Fig. 1). This geometry aims at maximizing the capacitive coupling between the STM tip and the RF antenna, which we discuss in more detail in Sect. III.B.

### B. Sample and tip preparation

A clean Ag(100) surface was prepared by repeated cycles of sputtering (for 10 min in an Ar atmosphere, sputtering current of 30 μA) and annealing (for 10 min, sample temperature of 800 K). Mg was evaporated from a resistively heated crucible held at a temperature of 653 K at a growth rate of 0.2 Å/min, during which the sample was kept at a temperature of 700 K in an $O_2$ atmosphere (1e-6 mBar). After 20 min of slow cool down in ultra-high vacuum (5e-10 mBar), the sample was inserted directly into the STM at 4.5 K [40-42]. In this way, we obtain rectangular MgO islands that are tens of nanometers in size (see Fig. 2a). The MgO thickness is characterized by point-contact measurements [42]. All single-atom experiments were performed on double-layer MgO, which is the most abundant thickness on our sample.

We deposited Fe and Ti atoms in-situ with an electron-beam evaporator with the sample kept below 10 K (see Fig. 2b). Fe atoms were deposited to prepare spin-polarized STM tips [20]. It is known that residual hydrogen gas in the vacuum chamber tends to hydrogenate the Ti atoms, forming TiH [43]. With respect to the O sublattice, Fe absorbs exclusively on top of O, whereas TiH can be found on an O-O bridge site ($TiH_B$) or atop an O atom ($TiH_O$). The different species on the surface are recognized by their specific $dI/dU$ spectra and their appearance in constant-current images [6,7,21,32]. $dI/dU$ spectra (see Fig. 2c) were obtained at constant-height with a (root-mean-square) bias-voltage modulation of 1.5 mV at 971 Hz.

The STM tip was made of a chemically etched W wire that we indented into the clean Ag substrate until an atomically sharp STM-tip apex was obtained. Spin-polarized STM tips were prepared by repeatedly picking up of Fe atoms from the surface (on the order of 10); the resulting spin polarization was confirmed by conductivity spectra on $TiH_O$ showing a pronounced step around zero DC bias that is otherwise absent (see inset in Fig. 2c) [21].

### C. Excitation and detection of EPR

The excitation mechanism of EPR-STM is believed to rely on a piezo-electric coupling of the RF electric field inside the tunnel junction to the EPR species leading to a GHz mechanical oscillation in the inhomogeneous exchange field caused by the nearby magnetic STM tip. The resulting change in the transverse effective magnetic field induces transitions between the Zeeman states split by an external magnetic field [25,26]. Upon resonance, the average Zeeman-state population changes and the junction resistance oscillates at the driving frequency. The former is equivalent to a decrease of the longitudinal magnetization of the probed atom, which is sensed by a DC read-out current through the magnetic tip as a tunnel magneto resistance effect. The latter is due to the precession of the transverse magnetization, which is sensed by the homodyne current resulting from the mixing of the AC junction conductance and the driving RF voltage [24,44]. As outlined in more detail in Ref. [24], this results in an EPR line shape that contains symmetric (DC and homodyne detection) and asymmetric (only homodyne detection) components, which can be described by a Fano line shape [24,44].

Other excitation mechanisms have been proposed [25-28], including the RF magnetic field due to the tunneling and displacement currents associated to the RF electric field between tip and sample, a RF modulation of the tunnel barrier in a cotunneling picture, a spin-transfer torque induced by the RF tunnel current, or the modulation of the dipolar field between the magnetic tip and the probed atom. Such mechanisms are considered too weak to be effective; however, more work is required in order to quantify and disentangle one from the other.

In our experiments, we modulate the amplitude of the microwave excitation voltage at 971 Hz with a square wave and a modulation depth of 100%, and detect the EPR-induced change of the tunneling current using a lock-in amplifier (LIA). Alternatively, we modulate the frequency of the microwave excitation voltage at 971 Hz with a square wave and a bandwidth of 32 MHz. The STM feedback parameters are the same for constant-current images, RF-transmission characterization, and EPR-STM measurements. An atom-tracking module was used during all measurements on single atoms, which gives a spatial averaging over a circle with radius of 10 pm set by the tracking-scheme's routine. In this work, the operational temperature of the STM was restricted to the liquid helium bath temperature of 4.5 K, which increased up to 5 K upon application of an RF signal at an RF-generator output power of $P_G \leq$ 30 dBm.

# III. COUPLING OF THE RF ANTENNA TO THE TUNNEL JUNCTION

## A. Transfer function

The coupling of the RF antenna to the tunnel junction is characterized following the scheme presented in Ref. [36] by measuring the RF transmission function $T_{RF}(f) = 10 \log U_{RF}/U_G$, where $f$ is the frequency, $U_{RF}$ is the RF-voltage amplitude, and $U_G$ is the output voltage of the RF generator. This definition of $T_{RF}$ has a more direct interpretation than the ones previously adopted in EPR-STM studies, which were based on either $U_{RF}$ relative to $P_G$ [36] or the RF power at the tunnel junction relative to $P_G$ assuming a tunneling-junction impedance of 50 Ω [34,37]. To simplify comparison with previous works, we also plot in Fig. 3a the transmission function $T_{RF,power} = 10 \log(50\,\Omega * U_{RF}^2/P_G)$ related to the latter definition. We obtain $U_G$ by converting the RF-generator output power $P_G$ using an impedance at its output of 50 Ω. Note that here all RF-voltage amplitudes and RF powers (modulated and unmodulated) refer to zero-to-peak values. Details about the calibration procedure are given in Appendix A. Notably, once calibrated, $T_{RF}$ remains approximately constant on a timescale of days, which we ascribe to the thermalization of the RF cables at fixed temperature points of the cryostat, whose temperatures change only little with the cryogenic's filling level (hold time of our system is about 100 h for liquid He).

Figure 3a shows $T_{RF}$ from 1 to 40 GHz. The detailed characterization of $T_{RF}$ allows us to analyze the different contributions to the transmission of the microwave excitation to the tunnel junction. This understanding is important for a future targeted optimization of $T_{RF}$.

First, we discuss the general features of $T_{RF}$. The data in Fig. 3a show that $T_{RF}$ is close to the estimated transmission function of the RF cables up to the antenna in a broad frequency range. The RF losses up to the RF antenna were estimated from the specifications of the cables in- and outside of the STM cryostat, and of all the connectors (see Appendix C for details). Further, we accounted for the different temperature stages in the cryostat and calculated a total RF-voltage loss of $(13 \pm 3)$ dB at 40 GHz from the RF generator to the antenna. For the estimated loss function, shown as the grey-shaded area in to Fig. 3a, we assume a linearly increasing dB-loss with frequency, which is mainly given by the coaxial cable. Upon comparison with the measured $T_{RF}$, we obtain the remarkable result that the RF-antenna can reach coupling efficiencies to the STM tunnel junction on the order of one.

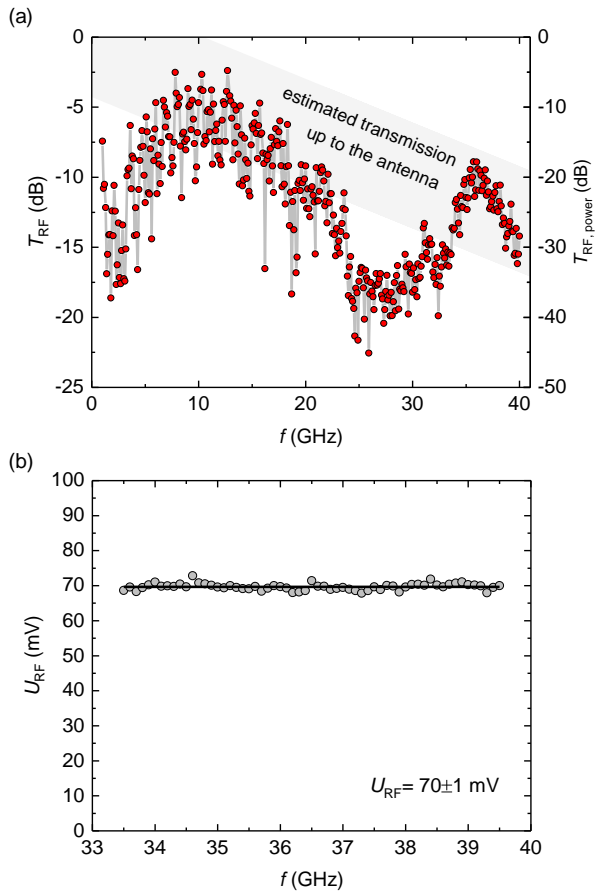

FIG. 3. (a) RF-transmission function $T_{RF}$ measured between 1 and 40 GHz (see Appendix A and B for details). Note that $T_{RF}$ is defined as the ratio of $U_{RF}$ to the voltage output by the RF-signal generator $U_G$. For comparison, also the RF-transmission function based on the ratio of RF powers $T_{RF,power}$ is shown. The grey-shaded area indicates the estimated RF transmission up to the RF antenna. (b) Compensating for $T_{RF}$ yields a constant $U_{RF}$ of 70 mV with a standard deviation of 1 mV. The solid line indicates the average $U_{RF}$. The entire calibration takes about 1 h. Settings: DC bias -90 mV, setpoint current 50 pA.

Upon closer inspection, $T_{RF}$ reveals two prominent oscillatory features: a fast oscillation with a period of several hundreds of MHz and an amplitude of about 5 dB, which we ascribe to standing waves along the entire length of the RF cabling, and a stronger modulation with an amplitude of about 15 dB (see the dip in $T_{RF}$ around 25 GHz in Fig. 3a). We ascribe this modulation to resonances of the RF antenna and its electromagnetic environment (compare with Fig. 1b), which includes the STM tip (length of about 5 mm), the gap between RF antenna and STM tip (also about 5 mm), and the surrounding metallic STM body (with dimensions in the centimeter range). This notion is further supported by a characterization of the entire RF-transmission line prior to installation. Measuring the RF losses with a vector network analyzer by inserting the open-ended flexible cable loosely into its input port, we observed a featureless transmission up to 40 GHz (not shown). This discussion highlights the importance of considering not only the RF cable itself but also its environment for an efficient RF coupling.

With the knowledge of $T_{RF}$, we can now compensate the RF losses in a broad spectral range by using an iterative optimization procedure, thereby obtaining a frequency-independent microwave excitation at the STM junction. In this way, we obtain, for instance, $U_{RF} = (70 \pm 1)$ mV between 33.5 and 39.5 GHz as shown in Fig. 3b. Note that the setting accuracy of $P_G$ of our RF generator is 0.01 dB, which implies that the theoretical accuracy of $U_{RF}$ upon compensation of $T_{RF}$ is given by $\Delta U_{RF} \geq U_{RF}^2(10^{0.001} - 1)/2 = 0.1$ mV. This shows that an optimal RF calibration would require a significantly longer averaging time of about 100 h (see Fig. 3b).

The extraordinary performance of our indirect RF-coupling scheme becomes apparent in comparison to previous reports, in which the RF excitation was fed directly through the STM-tip wiring into the tunnel junction [34,36]. Our $T_{RF}$ is on average 15 dB higher than the RF-transmission reported by Natterer et al. in the frequency range between 10 and 30 GHz (for comparison with our definition of $T_{RF}$, their transmission function was divided by 2) [34]. In the frequency range from 1 to 2 GHz, Hervé and coworkers [37] report an RF-transmission of about -10 dB (for comparison with our definition of $T_{RF}$, their transmission function was divided by 2), which is comparable to or better than our $T_{RF}$ in this low-frequency range. However the RF-transmission for higher frequencies was not reported in this work. For the frequency interval from 16 to 34 GHz, Paul et al. [36] find RF-transmissions ranging from -10 dB to -35 dB (for comparison with our definition of $T_{RF}$, their transmission function was reduced by 50 dB and divided by 2). In the same frequency window, our measured $T_{RF}$ (see Fig. 3a) ranges between -5 dB and -23 dB. This allows us to apply a more than ten times higher constant $U_{RF}$ in a frequency sweep for the same $P_G$, i.e., two orders of magnitude higher RF power at the tunnel junction.

**B. Equivalent circuit model**

Previous EPR-STM studies focused on characterizing $T_{RF}$ but did not analyze the coupling of the RF signal to the tunnel junction in detail [34,36,37]. In this regard, we now aim at analyzing the efficiency of the antenna coupling to the STM junction. For this purpose, we employ an equivalent circuit model (see Fig. 1c). Our model considers capacitive coupling from the RF antenna to the sample and to the STM tip with capacitances $C_S$ and $C_T$, respectively. The tunnel junction is modelled as a resistance $R_J$ (typically in the GΩ range) in parallel to a capacitance $C_J$ and, as a reference, we set the sample voltage to zero. This minimal model is sufficient to capture the impact of the electromagnetic surrounding on the RF coupling efficiency.

We measure $C_S$ and $C_T$ by applying a kHz voltage to the RF antenna while recording with a LIA the induced currents in the sample and the STM tip, respectively, giving for $C_S$ and $C_T$ values on the order of $10^{-14}$ F. In theory, the capacitance between a wire parallel to a plate is given by $C_S = 2\pi\varepsilon_0 l/\text{arcosh}(d/a)$, with the vacuum dielectric constant $\varepsilon_0$, the length $l$, the wire diameter $a$, and the distance $d$. The capacitance between two parallel wires is given by $C_T = C_S/2$. In our experiment, $l \approx d \approx 5$ mm (these are upper limits due to the angle between the antenna and the STM tip) and $a =$

0.1 mm, which yields calculated values for the capacitances of about $3 \cdot 10^{-14}$ F, in good agreement with the measured values. Moreover, reported values for $C_J$ range between $10^{-18}$ F and $10^{-15}$ F [45], implying that $C_{T,S} \gg C_J$.

The equivalent-circuit model corresponds to a voltage divider along the antenna/tip path, yielding

$$U_{RF} = U_1 \left( \frac{C_T + C_J}{C_T} + \frac{1}{2\pi i f R_J C_T} \right)^{-1}, \quad (1)$$

where $U_1$ is the voltage applied to the antenna. Note that $U_{RF}$ is independent of $C_S$; the latter, however, determines how the electrical current, and therefore the RF power, splits among the antenna/sample and antenna/tip paths. As $f$ is in the GHz range, the imaginary part in the denominator of Eq. 1 is negligible. This leads to the important conclusion that $U_{RF} \approx U_1$ if $C_T \gg C_J$, which explains the high coupling efficiency of the RF antenna observed experimentally (see Fig. 3a). Note that this situation is the long-wavelength/near-field analogue to the coupling of infrared radiation to a Whisker diode reported previously [46], where a thin metal tip acts as an efficient long-wire receiving antenna.

The electrostatic picture (i.e., the near-field coupling) employed above is justified if all the involved wavelengths (e.g. 30 cm at 1 GHz) are much larger than the typical length scales of our setup. More specifically, the Frauenhofer condition states that the far-field coupling starts dominating at distances from the antenna $L \geq 2fl^2/c$, where $c$ is the speed of light [47]. We find that $L \geq 7$ mm for the employed frequencies, which shows that the far-field coupling is not dominant in our experiment. Nevertheless, the wave nature of the RF signal might lead to deviations from the near-field picture outlined above, as apparent from the complex structure of the measured $T_{RF}$. Understanding of the latter requires a more sophisticated RF modelling, which is outside the scope of this study.

Knowing $C_T$ also allows us to estimate the strength of the antenna's RF magnetic field $B_A$ caused directly by the displacement current upon charging the antenna-tip capacitor (note that including the antenna-sample capacitor leads to minor corrections and, thus, to the same conclusions). According to Ampère's law, $B_A = \mu_0 f U_1 C_T / r$, where $r$ is the distance from the axis of the antenna-tip capacitor and $\mu_0$ is the vacuum permeability. With $f = 40$ GHz, $U_1 = 1$ V, and $r \approx 5$ mm, we find that $B_A \approx 10^{-4}$ mT at the STM junction, which results in a Rabi rate $\Omega = g\mu_B B_A / 2\hbar \approx 10^4$ Hz with the g-value for TiH of $g = 2$ [21], the reduced Planck constant $\hbar$ and the Bohr magneton $\mu_B$ [14]. According to Ref. [24], the maximum change in tunnel current sensed by the DC tunnel current is given by $\Delta I_{DC} \approx I_{DC} a_{TMR} (\Omega T_s)^2 / [1 + (\Omega T_s)^2]$, where the homodyne contribution to the current is neglected. Here, $a_{TMR}$ is the tunnel-magneto-resistance efficiency and $T_s$ is the spin life time, where equal longitudinal and transverse lifetimes are assumed. We estimate an upper bound for $\Delta I_{DC}$ by setting $I_{DC} = 10$ pA, $T_s \approx 100$ ns [24], and $a_{TMR} = 1$, and find $\Delta I_{DC} \approx 1$ aA, which is far below the detection limit of 10 fA in our setup. Therefore, $B_A$ cannot be the EPR driving source.

### C. Best frequency window for EPR-STM at 4 K

The spectral resolution, spin polarization, and sensitivity of EPR generally increase with increasing frequency or the associated static magnetic field. The best frequency window for EPR-STM is dictated by the following considerations. For a larger frequency $f$, the external magnetic field $B_{ext}$ has to increase in order to match the resonance condition. The larger $B_{ext}$ leads to a higher thermal population asymmetry $\propto \tanh[hf/2k_B T]$ between the Zeeman-split states (with the Planck constant $h$ and the Boltzmann constant $k_B$). Thus, a higher frequency favors a larger EPR signal. This reasoning is further supported by the additional increase of the STM-tip spin polarization with larger $B_{ext}$. In combination with the experimental observation that the EPR-signal amplitude $A$ scales linearly with $U_{RF}$ (see below), we find

$$A(f) \propto \tanh(hf/2k_B T) U_{RF}(f). \quad (2)$$

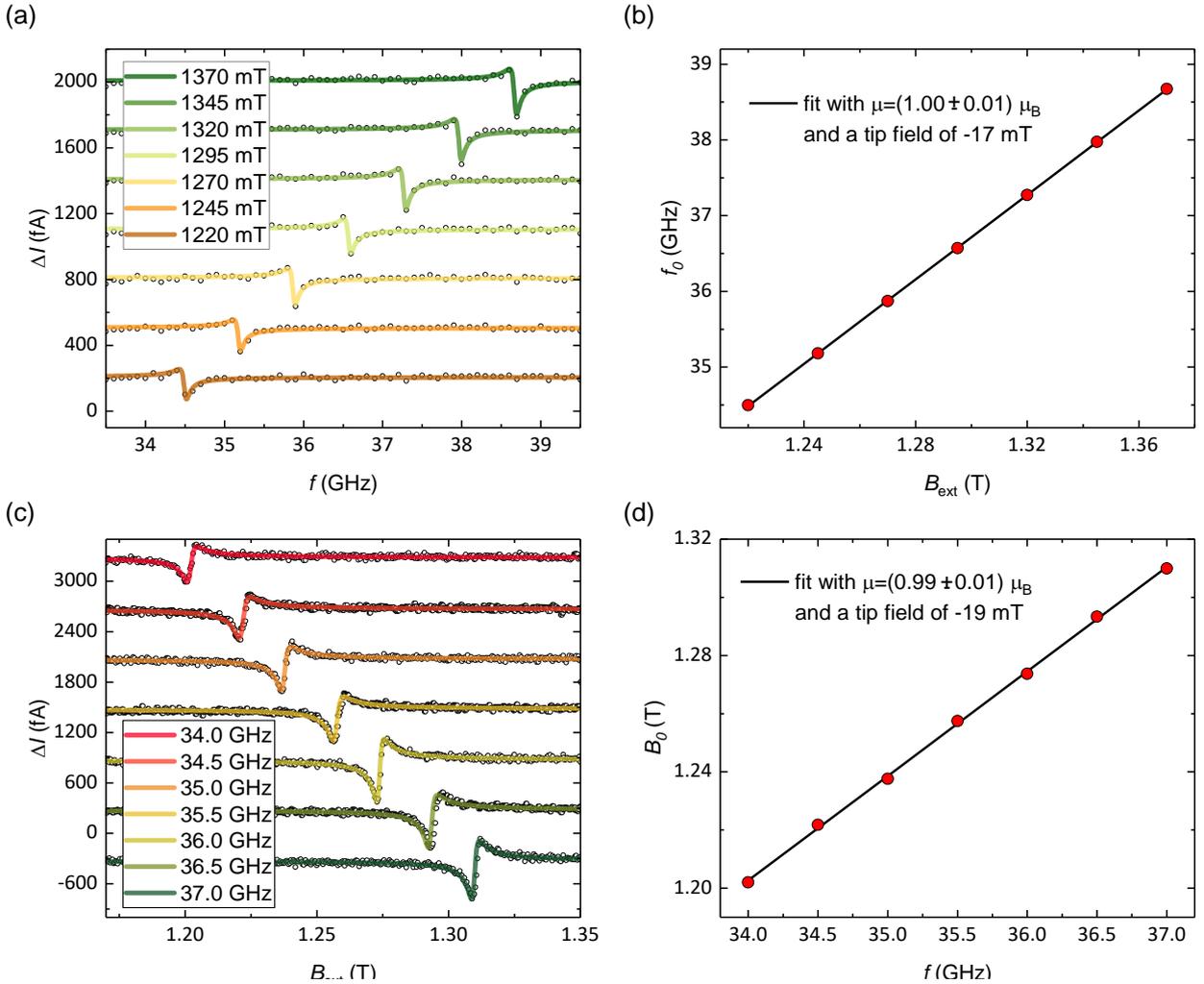

FIG. 4. EPR of a single hydrogenated Ti atom (TiH$_B$) in frequency- and magnetic-field-sweep mode. (a) Frequency sweep for different static external magnetic fields at a constant RF-voltage amplitude $U_{RF} = 70$ mV. Curves are offset in ascending order of the magnetic field for better visibility. (b) Fits by a Fano line shape of the data in (a) (see main text) allows to extract the resonance frequencies vs the applied magnetic field. A linear fit yields a magnetic moment for TiH$_B$ of $1.00 \pm 0.01$ $\mu_B$. (c) Magnetic-field sweeps for different frequencies at a constant RF-voltage amplitude of 150 mV. Curves are offset in descending order of the frequency for better visibility. (d) Same as (b) but fitting the data shown in (c) yielding the same magnetic moment for the same atom. All data were recorded with an amplitude-modulated RF voltage. The acquisition time for the data presented in (a) and (c) was about 10 min for a single spectrum. Settings: DC bias 100 mV, setpoint current 20 pA.

From this, we derive that (see Fig. 3a) the best frequency window for EPR-STM for our $T_{RF}$ is located above 30 GHz. Note that the minor raise in temperature of the STM body with applied microwave power has been neglected in Eq. 2.

To summarize the RF characterization, we find that, in general, higher frequencies are particularly suited for EPR-STM (see Eq. 2) and, since our $T_{RF}$ performs well above 30 GHz, it is exactly this previously unexplored frequency window from 30 to 40 GHz which is best suited for our experiments (see Fig. 3a).

## IV. EPR OF SINGLE HYDROGENATED TI ATOMS

Knowing the RF excitation precisely, we now describe and compare different measurements of the EPR of single magnetic atoms on a surface. To record an EPR spectrum, the STM tip was positioned above an isolated TiH$_B$ atom, at a distance larger than 2 nm from other magnetic species in order to minimize interactions.

In the following, we used two different schemes for EPR sweeps: In a magnetic-field sweep

(MFS), $f$ is constant, whereas in a frequency sweep (FS), $B_\text{ext}$ is constant. For the latter, we compensate $T_\text{RF}$ (see Figs. 3a and b) to avoid spurious signals. Note that a MFS requires a sufficiently high mechanical stability of the STM during a ramp of $B_\text{ext}$.

In the experiment, we measure the change in DC current $\Delta I$ (peak-to-peak current) induced by the modulated microwave excitation, which is obtained from the detected lock-in voltage $U_\text{LIA}$ through division by the gain of the transimpedance amplifier (1e9 V/A) and multiplication by $\pi/\sqrt{2}$. The latter accounts for the square RF-power modulation (with sine demodulation) and for the peak-to-peak value of $\Delta I$. The FS data were recorded with an averaging time of 80 ms per frequency point (time constant of 20 ms at the LIA) and an overall averaging over ten FSs. The MFS data were acquired with a 200 ms running-average time (time constant of 50 ms at the LIA) at a magnetic-field sweep rate of 0.7 mT/s without any further averaging. These averaging schemes where chosen in order to obtain for MFS and FS a similar acquisition time of about 10 min per spectrum for the data presented in Fig. 4 while maximizing the respective RF excitation. We note that the first 10 s of each MFS are subject to strong noise due to a relative motion between STM tip and sample, which is compensated for by the atom-tracking module.

Figure 4 compares side by side the EPR spectra obtained in FS and MFS mode on the same TiH$_B$ complex with the same STM microtip. In both modes, a resonant feature clearly evolves by changing either $B_\text{ext}$ (see Fig. 4a) or $f$ (see Fig. 4c). To gain more insight into the MFS and FS spectra, we fit the EPR signal with a Fano function (cf. Refs. [24,44] and Sect. II.C) given by

$$\Delta I(\varepsilon) = A \frac{(q\Gamma/2+\varepsilon)^2}{(\Gamma/2)^2+\varepsilon^2} + \delta, \qquad (3)$$

with the amplitude $A$, the offset $\delta$, the Fano factor $q$, and the line width $\Gamma$. Here, $\varepsilon$ is the magnetic field or frequency relative to the resonance positions $B_0$ or $f_0$, respectively, given by $\varepsilon = f - f_0$ in the FS mode and by $\varepsilon = B_0 - B_\text{ext}$ in the MFS mode. This definition takes into account that FS and MFS are inverted along the x-axis, i.e., by going higher in frequency at constant $B_\text{ext}$, the resonance is approached from the low-energy side, whereas for the MFS, this situation is inverted. As seen in Figs. 4 a and c, the EPR spectra are well described by Eq. 3 and the fit parameters contain valuable information that will be discussed in the following: First, for both measurement schemes, $A$ increases with $B_0$ and $f_0$, respectively (see Figs. 4 a and c), which we ascribe to an increased thermal population asymmetry between the two Zeeman states (see Eq. 2) and an increase in STM-tip spin polarization. We find values of $A$ about twice as high for the MFS than for the FS, which is attributed to the doubled RF-voltage amplitude (70 mV for FS and 150 mV for MFS) and indicates a dominating homodyne EPR-detection mechanism [24,33]. Second, from the fit we find the values $\Gamma = 90$ MHz (80 MHz) and $q = 0.6$ (0.7) for the MFS (FS), respectively. These small variations in $\Gamma$ and $q$ are expected since a different $U_\text{RF}$ was used for the MFS and FS measurements (see Fig. 4) [24]. The EPR line shapes appear similar to those reported in Ref. [32] although $q$ is not explicitly given therein.

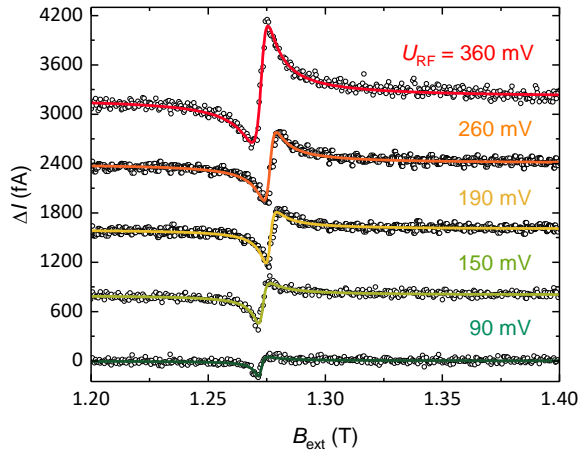

FIG. 5. EPR spectra of TiH$_B$ versus the RF-voltage amplitude. For better visibility, curves are vertically offset by 800 fA with respect to each other. Settings: frequency 36 GHz, DC bias 100 mV, setpoint current 20 pA. All data were recorded with an amplitude-modulated RF voltage.

Remarkably, a line width $\Gamma$ on the order of 100 MHz corresponds to an energy resolution better than 1 µeV even for temperatures as high as 5

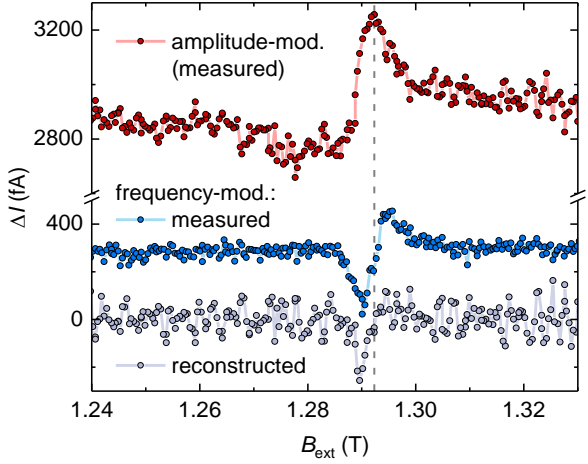

FIG. 6. Comparison of amplitude modulation (AM, red) and frequency modulation (FM, blue) schemes for EPR of TiH$_B$ recorded with magnetic-field sweeps. Settings: frequency 36 GHz, RF-voltage amplitude 260 mV, DC bias 100 mV, setpoint current 30 pA). AM settings: square-wave modulation, depth of 100%. FM settings: square-wave modulation, modulation bandwidth of 32 MHz.

K. This resolution is about three orders of magnitude below the thermal limit, which clearly demonstrates the advantage of EPR-STM over conventional scanning tunneling spectroscopy in terms of resolving magnetic excitations. An important result is that $\Gamma$ is similar to that reported for TiH$_B$ species measured at 1 K [24,34], indicating that the read-out process (i.e., the interaction with tunneling electrons) is still limiting the spin life time instead of the intrinsic temperature-dependent life time. We additionally verified that the relatively high RF-power levels that we deliver to the tunnel junction do not broaden the EPR spectra by local heating. For this purpose, the AM depth was varied in order to alter the average local RF-induced heating, which, however, did not influence $\Gamma$. These observations have significant implications as they show that, for the studied system, the energy resolution is not limited by temperature but only by the measurement process itself.

Regarding the resonance positions, linear fits of $f_0(B_{ext})$ and $B_0(f)$ (see Figs. 4 b and d) consistently yield a magnetic moment of $1.00 \pm 0.01$ $\mu_B$ for TiH$_B$, in accordance with previous DFT calculations [21] and measurements [21,34]. However, we find deviations from the magnetic moment reported in Ref. [24] of 0.9 $\mu_B$ for TiH$_B$. Bae et al. [24] argue that this change in measured moment arises from a finite angle between $B_{ext}$ and the tip magnetic field experienced by the atom on the surface, which is experimentally difficult to access [26]. Additionally, an STM-tip magnetic field of about 20 mT is found from the intercepts of the fits in Figs. 4 b and d, which is consistent with previous results [21]. Hence, we infer that MFS and FS provide equivalent results for measurements of EPR-STM.

Note that we found additionally broadened EPR peaks for a minority of the investigated TiH$_B$ complexes, which we interpret as indications of hyperfine interaction as reported in Ref. [32]. However, we could not yet resolve separate hyperfine-split EPR peaks even by reducing the setpoint current and $U_{RF}$. A likely limiting factor here is that we observe an additional extrinsic broadening of the EPR peaks, which is on the order of the hyperfine splitting of about 40 MHz for TiH$_B$ [32]. This additional broadening is attributed to the movement of the STM tip due to vibrations and the atom-tracking routine, which leads to variations of the exchange field between the STM tip and the magnetic atoms, as well as to the higher temperature of our setup.

### A. Maximizing the EPR excitation in magnetic-field sweeps

Next, we demonstrate one of the advantages of the MFS mode in combination with the RF-antenna coupling, which is the possibility to apply stronger EPR excitations. Accordingly, Fig. 5 presents MFS data for $U_{RF}$ from 90 to 360 mV at $f = 36$ GHz (with all other settings kept constant), which strongly affects the EPR signal. $A$ is found to increase linearly with $U_{RF}$ without saturation, showing how the MFS mode can increase the signal-to-noise ratio of EPR spectra by orders of magnitude. This result might be related to a dominating homodyne detection (cf. Sect. II.C) which was also reported previously for EPR of TiH$_B$ with $T \leq$ 1.2 K and $U_{RF} \leq 60$ mV [21,24,48].

### B. Frequency-modulated EPR-STM

All data presented up to here and all previously reported EPR-STM studies relied on amplitude modulation (AM) of the microwave excitation. However, for certain experimental conditions, it can be of advantage to use a different modulation scheme: If the $I(U)$ curve shows a strong nonlinearity in the range $U_{DC} \pm U_{RF}$, the AM leads to a large offset caused by the RF rectification (see Eq. S2 in the Appendix), which can result in an increased noise level by limiting the gain settings of the transimpedance amplifier and the LIA. To circumvent this issue, we introduce here EPR-STM performed by frequency modulation (FM) of $U_{RF}$ [14]. Figure 6 compares the EPR spectra recorded on the same TiH$_B$ atom with AM and FM using the same acquisition time (note that a different STM microtip was used compared to Figs. 4 and 5). Interestingly, the measured FM EPR spectrum $\Delta I_{FM}$ is similar to a derivative of the AM EPR spectrum $\Delta I_{AM}$ (see Fig. 6), as expected from classical EPR experiments [14], and a reconstruction of $\Delta I_{FM}$ is possible:

$$\Delta I_{FM}(x) \propto \Delta I_{AM}(x) - \Delta I_{AM}(x + \Lambda),$$

where $\Lambda$ is given by the FM bandwidth (32 MHz). The resulting reconstructed spectrum agrees well with the measured $\Delta I_{FM}$ (see Fig. 6). This demonstrates that AM and FM modes contain the same information and can thus be deliberately chosen according to the experimental requirements. We note that the signal-to-noise ratio in the FM mode could be enhanced by choosing a modulation bandwidth on the order of the EPR peak width (about 140 MHz for the data presented in Fig. 6), which, however, was not possible due to limitations of the modulation bandwidth of our RF generator.

## V. CONCLUSIONS

In summary, we demonstrated EPR measurements on single atoms driven by an RF-antenna with strong coupling efficiency to the STM junction and an energy resolution of about 1 μeV at 4-5 K. This approach allows for applying high RF-voltage amplitudes across the tunnel junction over a broad spectral range ($U_{RF}$ reaches 350 mV even above 30 GHz) and facilitates the implementation of EPR capabilities into standard 4 K STM.

Comparing the MFS and FS modes (see Fig. 4), we conclude that the MFS mode presents several advantages: First, it saves time by not requiring to characterize $T_{RF}$ in detail (about 1 h in our case), which might be of special importance if $T_{RF}$ changes on short time scales. Second, the MFS mode allows for selecting frequencies with good RF transmission, boosting the EPR-signal amplitude significantly (see Fig. 5).

We further demonstrated the possibility of performing EPR-STM by modulating the RF voltage in frequency, instead of amplitude. The FM mode is well established in classical EPR studies [14]. In EPR-STM, FM is of advantage compared to AM due to its weaker dependence on nonresonant background signals caused by heat modulation and its smaller sensitivity to characteristic features of $I(U)$ in the range $U_{DC} \pm U_{RF}$. Thus, by a proper choice of $f$, the FM mode can be of particular advantage in the spatial imaging of EPR of single atoms [48] as it is largely insensitive to local changes in the $I(U)$ curve within $U_{DC} \pm U_{RF}$, eliminating the need for background subtraction. Moreover, the gain of the STM preamplifier and the LIA can be increased to best match the amplitude of the resonance signal without reaching saturation, thus improving the signal-to-noise ratio for a given acquisition time. The precision in determining the position of the EPR signal can be increased further by choosing AM for asymmetric and FM for symmetric EPR spectra, respectively.

Our findings clearly show how the right choice of RF coupling and measurement mode can drastically enhance the signal-to-noise ratio of EPR-STM, allowing also for tailoring the EPR excitation to specific experimental requirements. Future work might aim at even higher driving frequencies to increase the EPR-signal amplitude further by reducing the thermal population of the excited state. A more sophisticated RF engineering of the entire cavity geometry might further maximize the RF-voltage amplitude, paving the way towards high-power pulsed EPR and coherent spin manipulation in STM [35].

# ACKNOWLEDGEMENTS

We acknowledge discussions with Prof. Katharina Franke that inspired the indirect RF coupling via the RF-antenna and Dr. Giovanni Boero, who commented on the measurements and provided early insight into EPR-STM. We also thank Dr. Pascal Leuchtmann for insightful discussions about microwave engineering. We acknowledge funding from the Swiss National Science Foundation, project # 200021_163225. L.P. was supported by an ETH Postdoctoral Fellowship (FEL-42 13-2).

# APPENDIX

## A. Calibration of the RF voltage at the tunnel junction

STM is usually performed in the low-frequency domain due to the small amplitude of the tunneling current (~pA), whose detection requires a low-frequency cut-off of the transimpedance amplifier (see Fig. 1c in the main text). Therefore, it is important to know how the presence of an RF voltage can influence STM measurements. In the experiment, an RF voltage at a frequency $f$ in the GHz range with amplitude $U_{\text{RF}}$ is added to the DC bias voltage $U_{\text{DC}}$ yielding a total voltage of

$$U = U_{\text{DC}} + U_{\text{RF}}\sin(2\pi ft). \quad \text{(S1)}$$

In STM, as in any nonlinear circuit, a second- and any higher even-order nonlinearity in the current-voltage characteristic $I(U)$ gives rise to an additional DC current upon rectification of the RF voltage in the resulting tunnel current $I(U)$ [49]. This can be readily understood by considering a Taylor expansion of $I(U)$ around $U_{\text{DC}}$ up to second order, which gives

$$I(U) = I(U_{\text{DC}}) + \left.\frac{dI}{dU}\right|_{U=U_{\text{DC}}} U_{\text{RF}}\sin(2\pi ft) + \frac{1}{2}\left.\frac{d^2I}{dU^2}\right|_{U=U_{\text{DC}}} [U_{\text{RF}}\sin(2\pi ft)]^2.$$

Considering that the transimpedance amplifier (see Fig. 1c in the main text) cannot detect a current oscillating far above its cut-off frequency of several kHz, we find

$$I(U) = I(U_{\text{DC}}) + \frac{1}{4}\left.\frac{d^2I}{dU^2}\right|_{U=U_{\text{DC}}} U_{\text{RF}}^2. \quad \text{(S2)}$$

Thus, a RF voltage causes an additional DC current, which, to lowest order, is proportional to $U_{\text{RF}}^2$ and to the second-order nonlinearity of the $I(U)$ characteristic. As a consequence, a $dI/dU$ spectrum recorded at finite $U_{\text{RF}}$ is also altered [49]. However, the Taylor-expansion approach is only valid if $U_{\text{RF}} \ll U_{\text{DC}}$. In the more general case of arbitrary $U_{\text{RF}}$ and for typical EPR-STM experimental conditions, the altered $I(U)$ or $dI/dU$ curve is determined by a convolution of the corresponding characteristic in the absence of RF voltage with the arcsine-distribution function carrying the information on $U_{\text{RF}}$ [36,50]. Here, the arcsine-distribution function describes the probability of the voltage taking a specific value in the interval $U_{DC} \pm U_{RF}$ over one oscillation period $1/f$.

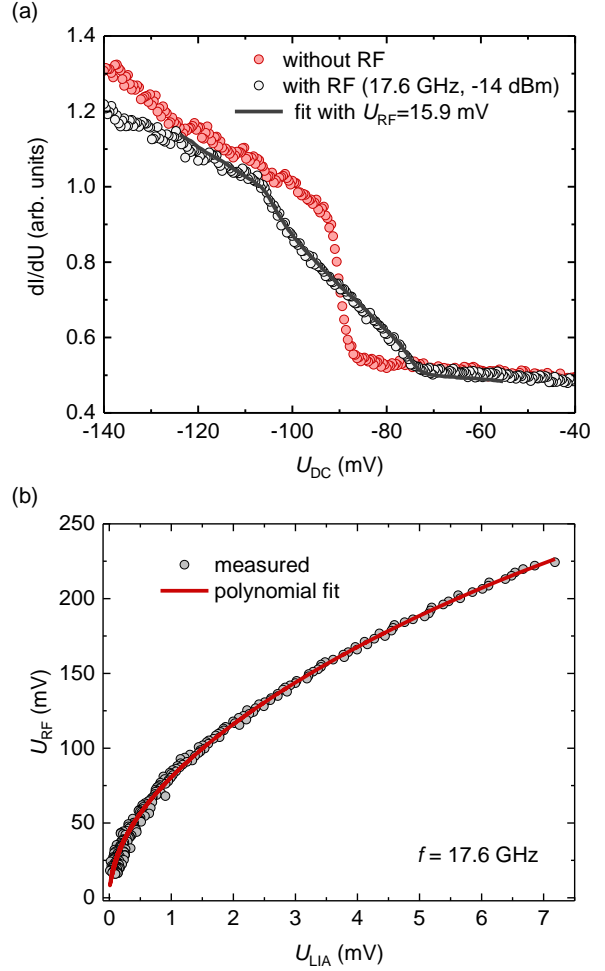

FIG. 7. Characterization of the RF-transmission function $T_{\text{RF}}$ to generate a frequency sweep at a constant RF-voltage amplitude $U_{\text{RF}}$ at the tunnel junction. (a) $dI/dU$ spectrum recorded on TiH$_\text{O}$ (current feedback opened at -40 mV, setpoint current 50 pA) with and without continuous-wave RF voltage. A fit of the broadened spectrum yields $U_{\text{RF}} = 15.9$ mV at the frequency $f = 17.6$ GHz with an RF-generator output power $P_{\text{G}} = -14$ dBm (see main text for details). (b) A sweep of $P_{\text{G}}$ (amplitude-modulated) at $f = 17.6$ GHz allows mapping of the detected lock-in-amplifier voltage $U_{\text{LIA}}$ to $U_{\text{RF}}(U_{\text{LIA}})$. For this purpose, we rescale $U_{\text{RF}}(U_{\text{LIA}})$ by the result from (a) and fit the curve with a polynomial. Settings for (b): DC bias -90 mV, setpoint current 50 pA.

## B. Measurement of the transfer function

In detail, $T_{RF}$ is characterized in three steps [36]:

First, we calibrate $U_{RF}$ for one specific pair of the RF-generator output power $P_G$ and frequency $f$. For that, we measure the $dI/dU$ spectrum of TiH$_O$, which has a strong nonlinearity due to a vibrational inelastic excitation at around -90 mV (see Fig. 2 in the main text). Upon application of $P_G = -14$ dBm (unmodulated) at $f = 17.6$ GHz, the $dI/dU$ spectrum is broadened (see Eq. S2 and Fig. 7a). This RF-voltage-induced broadening is reproduced by convoluting the $dI/dU$ spectrum measured in the absence of RF excitation with an arcsine-distribution function, fitting $U_{RF}$ in order to match the experimental broadening. Figure 3a shows a comparison of the $dI/dU$ spectra measured with $U_{RF} = 0$ and $U_{RF} = 15.9$ mV. Thus, we know $U_{RF}$ for one pair of $f$ and $P_G$.

Second, we sweep $P_G$ at $f = 17.6$ GHz. To maximize the rectified voltage (see Eq. S2), $U_{DC}$ is set to the steepest part of the $dI/dU$ curve, i.e., to -90 mV (cf. Fig. 2 in the main text). We modulate the RF signal in amplitude and record the RF-induced rectified current by the first harmonic in-phase signal $U_{LIA}$ measured by a LIA behind the transimpedance amplifier. Upon rescaling this RF-power sweep by the calibrated $U_{RF} = 15.9$ mV at $P_G = -14$ dBm and $f = 17.6$ GHz (see above and Fig. 7a), converting power to voltage, and taking the inverse function of $U_{LIA}(U_{RF})$, we can establish a polynomial relationship (of 4$^{th}$ order with zero offset) between the measured $U_{LIA}$ and $U_{RF}$, i.e., an analytic relation $U_{RF}(U_{LIA})$ (see Fig. 7b).

Third, $f$ is swept at $P_G = 9$ dBm while we record $U_{LIA}$. Using $U_{RF}(U_{LIA})$ finally allows us to determine $T_{RF}$ from 1 to 40 GHz (see Fig. 3a in the main text).

A frequency-independent RF-rectification efficiency is assumed for this calibration, which is justified for the given frequency range [49].

## C. Estimate of the RF-line losses

For the semi-rigid coaxial cable (see Sect. II.A for details) the manufacturer provides information on the attenuation for the frequencies 0.5, 1, 5, 10 and 20 GHz at the temperatures of 300 and 4 K. In our analysis we calculated the power attenuation of the cables with an impedance of 50 Ω using the formula

$$\alpha[\text{dB/m}] = 1.7372 \cdot \sqrt{f_{GHz}} \left( \frac{\sqrt{\mu_{D_c} \rho_{D_c}}}{D_c} + \frac{\sqrt{\mu_{D_i} \rho_{D_i}}}{D_i} \right) + 92.0216 \cdot \sqrt{\varepsilon_r} \tan \delta \cdot f_{GHz},$$

where $\varepsilon_r$ is the relative permittivity of the dielectric, $\rho_{D_i}$ and $\rho_{D_c}$ the resistivity of the inner and outer conductor, respectively, $D_i$ the inner diameter of the outer conductor, $D_c$ the diameter of the center conductor, $\tan \delta$ the dielectric loss tangent and $f_{GHz}$ the frequency in GHz. The relative dielectric constant of the dielectric material PTFE is $\varepsilon_r = 2.02$. PTFE also has a very low loss tangent with a typical value of $\tan \delta = 4 \cdot 10^{-4}$, which decreases by a factor of 2-3 from 300 K down to 4 K [51,52]. Here, we used $\tan \delta = 2 \cdot 10^{-4}$ at 4 K. The relative magnetic permeabilities of the inner and outer conductors $\mu_d$ and $\mu_D$, respectively, were set to 1 for the employed frequency range. The values for $d$ and $D$ are tabulated in the data sheet of the cables. The resistivities of the silver-plated CuBe inner conductor and CuBe outer conductors have been estimated as $\rho_{D_c}(300 \text{ K}) = 1.55 \cdot 10^{-8}$ Ωm, $\rho_{D_c}(4 \text{ K}) = 5.0 \cdot 10^{-11}$ Ωm, $\rho_{D_i}(300 \text{ K}) = 8.15 \cdot 10^{-8}$ Ωm, and $\rho_{D_i}(4 \text{ K}) = 4.7 \cdot 10^{-8}$ Ωm following published data on low-temperature materials properties [53]. With these values we are able to reproduce the tabulated losses from the manufacturer within an error of 0.2 dB, while we can calculate the attenuation at an arbitrary frequency value. At 40 GHz, we calculate the power attenuation of the coaxial cable to be 10.2 dB/m at 300 K and 3.9 dB/m at 4 K.

Below about 40 K the residual resistivity of most metals does not change anymore due to the residual resistivity from material imperfections, i.e., the RF losses are expected to saturate at a minimum value below these temperatures [53]. For the semi-rigid cable, we have a section of 1 m connecting the feedthrough flange at 300 K with the radiation shield at about 40 K. For this section we used the average attenuation between 300 K and 4 K. For the last section of 1.5 m we used the attenuation for 4 K. Hence, the estimated power attenuation of the entire semi-rigid coaxial cable is 13 dB at 40 GHz.

Further components to consider in the attenuation analysis are the DC block at the output of the generator (1 dB flat), the 1.5 m long coaxial cable from the generator to the feedthrough flange (4 dB flat), the feedthrough flange (1 dB flat), a K-type to SMPM adapter (1 dB flat) at the end of the semi-rigid cable and the 30 cm long flexible coaxial cable. The latter has been measured at room temperature for a length of 1 m with two SMP connectors at its ends and the power loss was found to be 45 dB at 40 GHz. The estimated loss at 4 K (assuming half the losses compared to 300 K as found for the semi-rigid cable) with one connector only and a length of 30 cm is 7 dB. This adds 14 dB at 40 GHz to the 13 dB found for the semi-rigid cable. The total attenuation of 27 dB needs to be divided by 2 to be compared with the losses defined by the voltage ratio in the manuscript. Hence, we expect a voltage attenuation of $13 \pm 3$ dB as quoted in the manuscript, where the uncertainty mainly related to the losses in the flexible coaxial cable and the K-type-to-SMPM adapter.

References

[1] G. Binnig, H. Rohrer, C. Gerber, and E. Weibel, Surface studies by scanning tunneling microscopy, Phys Rev Lett **49**, 57 (1982).
[2] A. J. Heinrich, J. A. Gupta, C. P. Lutz, and D. M. Eigler, Single-atom spin-flip spectroscopy, Science **306**, 466 (2004).
[3] M. Ternes, A. J. Heinrich, and W. D. Schneider, Spectroscopic manifestations of the Kondo effect on single adatoms, J Phys-Condens Mat **21**, 053001 (2009).
[4] R. Wiesendanger, H.-J. Güntherodt, G. Güntherodt, R. Gambino, and R. Ruf, Observation of vacuum tunneling of spin-polarized electrons with the scanning tunneling microscope, Phys Rev Lett **65**, 247 (1990).
[5] T. Choi, Studies of single atom magnets via scanning tunneling microscopy, J Magn Magn Mater **481**, 150 (2019).
[6] S. Baumann *et al.*, Origin of Perpendicular Magnetic Anisotropy and Large Orbital Moment in Fe Atoms on MgO, Phys Rev Lett **115**, 237202 (2015).
[7] I. G. Rau *et al.*, Reaching the magnetic anisotropy limit of a 3d metal atom, Science **344**, 988 (2014).
[8] H. Brune and P. Gambardella, Magnetism of individual atoms adsorbed on surfaces, Surf Sci **603**, 1812 (2009).
[9] H. F. Hess, R. B. Robinson, and J. V. Waszczak, Stm Spectroscopy of Vortex Cores and the Flux Lattice, Physica B **169**, 422 (1991).
[10] H. Fukuyama, H. Tan, T. Handa, T. Kumakura, and M. Morishita, Construction of an ultra-low temperature scanning tunneling microscope, Czech J Phys **46**, 2847 (1996).
[11] S. Pan, E. W. Hudson, and J. Davis, 3He refrigerator based very low temperature scanning tunneling microscope, Rev Sci Instrum **70**, 1459 (1999).
[12] Y. J. Song, A. F. Otte, V. Shvarts, Z. Zhao, Y. Kuk, S. R. Blankenship, A. Band, F. M. Hess, and J. A. Stroscio, Invited review article: A 10 mK scanning probe microscopy facility, Rev Sci Instrum **81**, 121101 (2010).
[13] W. Tao *et al.*, A low-temperature scanning tunneling microscope capable of microscopy and spectroscopy in a Bitter magnet at up to 34 T, Rev Sci Instrum **88**, 093706 (2017).
[14] S. A. Al'Tshuler and B. M. Kozyrev, *Electron paramagnetic resonance* (Academic Press, 2013).
[15] M. Goswami, A. Chirila, C. Rebreyend, and B. de Bruin, EPR Spectroscopy as a Tool in Homogeneous Catalysis Research, Top Catal **58**, 719 (2015).
[16] D. Marsh, in *Membrane spectroscopy* (Springer, 1981), pp. 51.
[17] S. T. Liddle and J. van Slageren, Improving f-element single molecule magnets, Chemical Society Reviews **44**, 6655 (2015).
[18] Y. Manassen, R. Hamers, J. Demuth, and A. Castellano Jr, Direct observation of the precession of individual paramagnetic spins on oxidized silicon surfaces, Phys Rev Lett **62**, 2531 (1989).
[19] A. V. Balatsky, M. Nishijima, and Y. Manassen, Electron spin resonance-scanning tunneling microscopy, Advances in Physics **61**, 117 (2012).
[20] S. Baumann, W. Paul, T. Choi, C. P. Lutz, A. Ardavan, and A. J. Heinrich, Electron paramagnetic resonance of individual atoms on a surface, Science **350**, 417 (2015).
[21] K. Yang *et al.*, Engineering the Eigenstates of Coupled Spin-1/2 Atoms on a Surface, Phys Rev Lett **119**, 227206 (2017).
[22] K. Yang, P. Willke, Y. Bae, A. Ferron, J. L. Lado, A. Ardavan, J. Fernandez-Rossier, A. J. Heinrich, and C. P. Lutz, Electrically controlled nuclear polarization of individual atoms, Nat Nanotechnol **13**, 1120 (2018).
[23] R. Wiesendanger, Spin mapping at the nanoscale and atomic scale, Rev Mod Phys **81**, 1495 (2009).
[24] Y. Bae, K. Yang, P. Willke, T. Choi, A. J. Heinrich, and C. P. Lutz, Enhanced quantum coherence in exchange coupled spins via singlet-triplet transitions, Sci Adv **4**, eaau4159 (2018).
[25] J. L. Lado, A. Ferrón, and J. Fernández-Rossier, Exchange mechanism for electron paramagnetic resonance of individual adatoms, Phys Rev B **96**, 205420 (2017).
[26] K. Yang *et al.*, Tuning the Exchange Bias on a Single Atom from 1 mT to 10 T, Phys Rev Lett **122**, 227203 (2019).
[27] A. M. Shakirov, A. N. Rubtsov, and P. Ribeiro, Spin transfer torque induced paramagnetic resonance, Phys Rev B **99**, 054434 (2019).
[28] J. R. Gálvez, C. Wolf, F. Delgado, and N. Lorente, Cotunneling mechanism for all-electrical electron spin resonance of single adsorbed atoms, Phys Rev B **100**, 035411 (2019).
[29] T. Choi, C. P. Lutz, and A. J. Heinrich, Studies of magnetic dipolar interaction between

individual atoms using ESR-STM, Curr Appl Phys **17**, 1513 (2017).

[30] T. Choi, W. Paul, S. Rolf-Pissarczyk, A. J. Macdonald, F. D. Natterer, K. Yang, P. Willke, C. P. Lutz, and A. J. Heinrich, Atomic-scale sensing of the magnetic dipolar field from single atoms, Nat Nanotechnol **12**, 420 (2017).

[31] F. D. Natterer, K. Yang, W. Paul, P. Willke, T. Y. Choi, T. Greber, A. J. Heinrich, and C. P. Lutz, Reading and writing single-atom magnets, Nature **543**, 226 (2017).

[32] P. Willke *et al.*, Hyperfine interaction of individual atoms on a surface, Science **362**, 336 (2018).

[33] P. Willke, W. Paul, F. D. Natterer, K. Yang, Y. Bae, T. Choi, J. Fernandez-Rossier, A. J. Heinrich, and C. P. Lutz, Probing quantum coherence in single-atom electron spin resonance, Sci Adv **4**, eaaq1543 (2018).

[34] F. D. Natterer, F. Patthey, T. Bilgeri, P. R. Forrester, N. Weiss, and H. Brune, Upgrade of a low-temperature scanning tunneling microscope for electron-spin resonance, Rev Sci Instrum **90**, 013706 (2019).

[35] T. D. Ladd, F. Jelezko, R. Laflamme, Y. Nakamura, C. Monroe, and J. L. O'Brien, Quantum computers, Nature **464**, 45 (2010).

[36] W. Paul, S. Baumann, C. P. Lutz, and A. J. Heinrich, Generation of constant-amplitude radio-frequency sweeps at a tunnel junction for spin resonance STM, Rev Sci Instrum **87**, 074703 (2016).

[37] M. Hervé, M. Peter, and W. Wulfhekel, High frequency transmission to a junction of a scanning tunneling microscope, Appl Phys Lett **107**, 093101 (2015).

[38] A. Roychowdhury, M. Dreyer, J. R. Anderson, C. J. Lobb, and F. C. Wellstood, Microwave Photon-Assisted Incoherent Cooper-Pair Tunneling in a Josephson STM, Phys Rev Appl **4**, 034011 (2015).

[39] G. P. Kochanski, Nonlinear Alternating-Current Tunneling Microscopy, Phys Rev Lett **62**, 2285 (1989).

[40] S. Schintke, S. Messerli, M. Pivetta, F. Patthey, L. Libioulle, M. Stengel, A. De Vita, and W.-D. Schneider, Insulator at the ultrathin limit: MgO on Ag (001), Phys Rev Lett **87**, 276801 (2001).

[41] S. Schintke and W. D. Schneider, Insulators at the ultrathin limit: electronic structure studied by scanning tunnelling microscopy and scanning tunnelling spectroscopy, J Phys-Condens Mat **16**, R49 (2004).

[42] W. Paul, K. Yang, S. Baumann, N. Romming, T. Choi, C. P. Lutz, and A. J. Heinrich, Control of the millisecond spin lifetime of an electrically probed atom, Nat Phys **13**, 403 (2017).

[43] F. Natterer, F. Patthey, and H. Brune, Quantifying residual hydrogen adsorption in low-temperature STMs, Surf Sci **615**, 80 (2013).

[44] U. Fano, Effects of configuration interaction on intensities and phase shifts, Phys Rev **124**, 1866 (1961).

[45] S. Kurokawa and A. Sakai, Gap dependence of the tip-sample capacitance, J Appl Phys **83**, 7416 (1998).

[46] L. Matarrese and K. Evenson, Improved coupling to infrared whisker diodes by use of antenna theory, Appl Phys Lett **17**, 8 (1970).

[47] C. A. Balanis, *Antenna theory: analysis and design* (John wiley & sons, 2016).

[48] P. Willke, K. Yang, Y. Bae, A. J. Heinrich, and C. P. Lutz, Magnetic resonance imaging of single atoms on a surface, Nat Phys, 1 (2019).

[49] H. Nguyen, P. Cutler, T. E. Feuchtwang, Z.-H. Huang, Y. Kuk, P. Silverman, A. Lucas, and T. E. Sullivan, Mechanisms of current rectification in an STM tunnel junction and the measurement of an operational tunneling time, IEEE Trans Electron Devices **36**, 2671 (1989).

[50] P. K. Tien and J. P. Gordon, Multiphoton Process Observed in Interaction of Microwave Fields with Tunneling between Superconductor Films, Phys Rev **129**, 647 (1963).

[51] M. V. Jacob, J. Mazierska, K. Leong, and J. Krupka, Microwave properties of low-loss polymers at cryogenic temperatures, IEEE Transactions on Microwave Theory and Techniques **50**, 474 (2002).

[52] J. Krupka, Measurements of the complex permittivity of low loss polymers at frequency range from 5 GHz to 50 GHz, IEEE Microwave and Wireless Components Letters **26**, 464 (2016).

[53] S. W. Van Sciver, in *Helium Cryogenics* (Springer, 2012), pp. 17.